\documentclass[aps,prd,twocolumn,showpacs,floatfix,preprintnumbers,amsmath,amssymb, notitlepage, useAMS, superscriptaddress, nofootinbib]{revtex4-1}

\usepackage{amsmath}
\usepackage{amssymb}
\usepackage{hhline}
\usepackage{graphicx}
\usepackage{color}
\input epsf
\usepackage{mathtools}
\usepackage{mathbbol}

\newcommand{\be}{\begin{equation}}
\newcommand{\ee}{\end{equation}}
\newcommand{\barr}{\begin{eqnarray}}
\newcommand{\earr}{\end{eqnarray}}

\newcommand{\beq}{\begin{equation}}
\newcommand{\eeq}{\end{equation}}

\newcommand{\lya}{Ly$\alpha$ }
\newcommand{\LYA}{\rm{Ly}\alpha}
\newcommand{\dtb}{\delta T_{21}}
\newcommand{\Dtb}{T_{21}}

\newcommand\ion[2]{\textrm{#1} \textsc{#2}}

\newcommand{\xhi}{x_{\ion{H}{i}}}

\newcommand{\rme}{\textrm{e}}

\newcommand{\aap}{Astron. Astrophys.}

\newcommand{\aj}{Astron. J.}

\newcommand{\mnras}{Mon. Not. R. Astron. Soc.}

\newcommand{\physrep}{Phys. Rep.}
\newcommand{\araa}{Annual Rev. Astron. Astrophys.}
\newcommand{\jcap}{J. Cosm. Astropart. Phys.}

\newcommand{\procspie}{Proc. SPIE}

\usepackage{color}

\begin{document}
\title{The ultimate frontier of 21cm cosmology}
\author{Patrick C. Breysse}
\email{pcbreysse@cita.utoronto.ca}
\affiliation{Canadian Institute for Theoretical Astrophysics, 60 St George St., Toronto, ON, M5S 3H8, Canada}
\affiliation{Department of Physics and Astronomy, Johns Hopkins University, Baltimore, MD 21218 USA}
\author{Yacine Ali-Ha\"imoud}
\affiliation{Center for Cosmology and Particle Physics, Department of Physics, New York University, NY 10003, New York, USA}
\author{Christopher M. Hirata}
\affiliation{Center for Cosmology and AstroParticle Physics, Department of Physics, The Ohio State University, 191 W Woodruff Ave, Columbus OH 43210, USA}

\label{firstpage}

\begin{abstract}
We present the most detailed computation to date of the 21-cm global signal and fluctuations at $z\gtrsim 500$.  Our calculations include a highly precise estimate of the Wouthuysen-Field (WF) effect and the first explicit calculation of the impact of free-free processes, the two dominant components of the signal at $z\gtrsim 800$.  We implement a new high-resolution \lya radiative transfer calculation, coupled to a state-of-the-art primordial recombination code. Using these tools, we find a  global signal from 21-cm processes alone of roughly 0.01mK at $z\sim1000$, slightly larger than it would be without the WF effect, but much weaker than previous estimates including this effect. We also find that this signal is swamped by a smooth $1-2$ mK signal due to free-free absorption at high redshift by the partially ionized gas along the line of sight. In addition, we estimate the amplitude of 21-cm fluctuations, of order $\sim 10^{-7}$ mK  at $z\sim1000$. Unfortunately, we find that due to the brightness of the low-frequency sky, these fluctuations will not be observable beyond $z\sim$ a few hundred by even extremely futuristic observations.  The 21 cm fluctuations are exponentially suppressed at higher redshifts by the large free-free optical depth, making this the ultimate upper redshift limit for 21-cm surveys.
\end{abstract}

\maketitle

\section{Introduction}
\label{sec:intro}
In the most distant reaches of the Universe, between the onset of cosmological recombination and the end of reionization, hydrogen atoms provide one of the only available tracers of large-scale structure.  In particular, the 21-cm hyperfine transition in neutral hydrogen has become recognized as a powerful probe of these so-called ``dark ages" \citep{Furlanetto2006,Morales2010,Pritchard2012,McQuinn2016}.  This transition is optically thin, so it is in principle possible to map the structure of the high-redshift Universe in three dimensions by measuring angular and spectral fluctuations in brightness.

Most theoretical and experimental efforts to date have been focused on the 21-cm signal from the epoch of reionization, the last large-scale phase transition in the Universe where the intergalactic medium was ionized by the first galaxies \cite{Madau1997}. Radio instruments such as the LOw Frequency ARay (LOFAR, \cite{vanHaarlem2013}), the Precision Array for Probing the Epoch of Reionization (PAPER, \cite{Ali2015}), and the Murchison Wide-field Array (MWA, \cite{Tingay2013}) have begun placing upper limits on the EoR signal in preparation for planned experiments like the under-construction Hydrogen Epoch of Reionization Array (HERA, \cite{DeBoer2016}), the Square Kilometer Array (SKA, \cite{Santos2015}), and the Dark Ages Radio Explorer (DARE, \cite{Burns2012}). Other projects, including the Canadian Hydrogen Intensity Mapping Experiment (CHIME, \cite{Bandura2014}) and the Hydrogen Intensity and Real-time Analysis eXperiment (HIRAX, \cite{Newburgh2016}), will target the remaining neutral hydrogen within galaxies after reionization.

While the detection of the 21-cm signal from reionization seems imminent, it depends sensitively on complex astrophysical processes, and its interpretation will be a theoretical challenge \cite{Cohen2016}. It is also in principle possible to go after the 21-cm signal from beyond reionization \citep{Loeb2004, Lewis2007}. Though atmospheric and astrophysical foreground emission pose a significant technical challenge to observations at these redshifts \citep{Brown1973,Manning2001}, the signal is expected to be much cleaner of astrophysical uncertainties, in particular at $z \gtrsim 20-30$, before the first stars have formed \citep{Barkana2016}. High-redshift 21-cm tomography promises to become the ultimate cosmological probe by opening a window into the earliest periods of structure formation. To cite only a few applications, it can shed light on the character of initial conditions through the measurement of small-scale perturbations \cite{Loeb2004, AliHaimoud2014}, statistical isotropy \cite{Shiraishi2016} and primordial non-gaussianity \cite{Cooray2006, Pillepich2007, Munoz2015}, and can probe the nature of dark matter \cite{Furlanetto2006b, Natarajan2009, Tashiro2014, Munoz2015b}.

The 21-cm signal is the contrast between the brightness temperature in the redshifted 21-cm line and the CMB temperature. This contrast is itself proportional to the difference between the spin temperature, characterizing the relative abundance of ground-state hydrogen in its two hyperfine substates, and the CMB temperature. This is set by three competing processes: absorption of and stimulated emissions by resonant CMB photons, collisions between hydrogen atoms, and spin-flipping Lyman-$\alpha$ scatterings. The latter process, the Wouthuysen-Field (WF) effect \citep{Wouthuysen1952,Field1958}, indirectly drives the spin temperature towards the gas temperature, as the Lyman-$\alpha$ color temperature near line center is expected to be very near equilibrium with the kinetic temperature of the gas. The standard lore is that the WF effect is only relevant at low redshifts, once the first galaxies have produced a significant Lyman-$\alpha$ background. Recently, however, Ref.~\cite{Fialkov2013} (hereafter FL13) pointed out that the WF effect actually dominates over other processes at $z \gtrsim 850$, due to the large Lyman-$\alpha$ background generated during the out-of-equilibrium cosmological recombination process \cite{Peebles1968, Zeldovich1969}. More interestingly, FL13 found a significant departure of the Lyman-$\alpha$ color temperature from the gas temperature at $z \gtrsim 500$; as a result, they found a much larger 21-cm signal at these redshifts than previously estimated, though the precise value depends on the recombination code they used \cite{AliHaimoud2011, Chluba2011}.

The goal of this work is to thoroughly examine the little-explored 21-cm signal from $z \gtrsim 500$. We improve upon the first study by FL13 in several ways. First, we develop a very-high-resolution \lya radiative transfer code to ensure that the physics of the WF effect is captured precisely. This code is coupled to the recombination code \textsc{HyRec} \cite{AliHaimoud2010b, AliHaimoud2011}, which very accurately predicts the high-redshift recombination and thermal history, as well as the damping wings of the \lya line. We find that the fractional difference between color and matter temperatures at $z \gtrsim 500$ is at most a few times $10^{-5}$, up to three orders of magnitude less than the deviation found by FL13. 

Secondly, we explicitly account for free-free emission and absorption, which become substantial as the free-electron abundance rises rapidly at $z \gtrsim 10^3$. They lead to a distortion to the global CMB spectrum proportional to a weighted average of the difference between matter and radiation temperature, of order $\sim 1$ mK. This distortion overwhelms that due to resonant interaction in the 21-cm line for frequencies $\nu \lesssim 2$ MHz. In other words, the \emph{global} distortion to the CMB spectrum at frequencies $\nu \lesssim 2$ MHz has little to do with the 21-cm transition, and is mostly determined by free-free processes.

We also consider the \emph{fluctuating} part of the signal. Given that the free-free contribution results from a broad integral along the line of sight, we expect it to be mostly smooth, and the fluctuations should indeed be dominated by the 21-cm signal.  We find perturbations on the order of $\sim 10^{-7}$ mK at $z\sim1000$. Beyond that redshift, the high free-free optical depth exponentially suppresses any fluctuations, implying that this is the most distant redshift at which a 21-cm signal can be observed, as anticipated in FL13.  We also make a rough estimate of the sensitivity necessary to detect these fluctuations, and find that even a futuristic experiment densely covering the far side of the Moon will be unable to make a detection.

This paper is organized as follows.  In Section \ref{sec:background},we briefly review the physics underlying the 21 cm signal.  We discuss the global signal around recombination in Section \ref{sec:globalsignal}, starting with a description of our high-precision \lya radiative transfer computation and going on to add in the effects of free-free absorption.  Section \ref{sec:fluctuations} studies the amplitude of 21 cm fluctuations at these redshifts.  We discuss our results in Section \ref{sec:discussion} and conclude in Section \ref{sec:conclusion}.  Appendix \ref{app:numerical} provides the details of our numerical implementation, followed by a discussion of convergence tests in Appendix \ref{app:convergence}.  We assume a flat $\Lambda$CDM cosmology throughout with parameters consistent with the latest measurements.  For simplicity, we take $k_B=1$ throughout.


\section{Background}
\label{sec:background}

\subsection{General considerations}

The observable for 21-cm surveys is the difference $\Dtb\equiv T_b-T_{\rm{CMB}}$ between the brightness temperature $T_b$ at a given frequency and the CMB temperature $T_{\rm{CMB}}$. An observed frequency $\nu_{\rm{obs}}$ maps to redshift $z \equiv \nu_{21}/\nu_{\rm obs} -1$, where $\nu_{21} \approx 1.4$ GHz is the frequency of the spin-flip transition. The brightness temperature contrast is given by
\be
\Dtb=\frac{T_S(z)-T_r(z)}{1+z}\left(1-e^{-\tau_\nu}\right)\approx\frac{T_S(z)-T_r(z)}{1+z}\tau_\nu,
\label{DeltaT}
\ee
The observed intensity depends on the difference between the radiation temperature $T_r=T_{\rm{CMB}}(1+z)$ and the spin temperature $T_S$, defined by
\be
\frac{n_1}{n_0}=3e^{-T_\star/T_S} \approx 3 \left(1 - \frac{T_*}{T_S} \right),
\ee
where $n_1$ and $n_0$ are the densities of atoms in the upper ($F = 1$) and lower ($F = 0$) hyperfine states, $T_\star \equiv h \nu_{21} =0.068$ K is the temperature corresponding to the energy difference between these levels, and the second approximation is always accurate since all relevant temperatures are significantly higher than $T_*$.

The optical depth $\tau_\nu$ of the 21-cm transition is given by
\be
\tau_\nu=\frac{3}{32\pi}\frac{hc^3A_{10}}{T_S\nu^2}\frac{x_{\ion{H}{i}}n_{\rm{H}}}{(1+z)(dv_\parallel/dr_\parallel)},
\label{tau21}
\ee
where $h$ is Planck's constant, $c$ is the speed of light, $n_{\rm H}$ is the number density of hydrogen atoms, $\xhi$ is the neutral fraction of those atoms, and $A_{10}$ is the Einstein-A coefficient for the 21-cm line.  The quantity $(dv_\parallel/dr_\parallel)$ is the derivative of the line-of-sight gas velocity along the photon's path, including both the bulk Hubble flow and peculiar velocities \citep{Morales2010}.  Combining Equations (\ref{DeltaT}) and (\ref{tau21}) yields, in the matter-domination era \cite{Furlanetto2006},
\barr
\Dtb \approx 9\ \textrm{mK}\ \xhi(1+\delta)(1+z)^{1/2}\left[1-\frac{T_r}{T_S}\right] \nonumber\\
\times \left[\frac{H(z)/(1+z)}{dv_\parallel/dr_\parallel}\right], \label{eq:T21-num}
\earr
where $\delta$ is the density contrast at the location observed and $H(z)$ is the Hubble parameter.

The spin temperature is determined by three competing processes. Interactions with CMB photons tend to bring the spin temperature in equilibrium with the radiation temperature at a given redshift. Collisions between hydrogen atoms pull the spin temperature instead towards the matter temperature $T_m$. The third process,  the WF effect, alters the spin temperature through interactions with \lya photons. Indeed, electric dipole selection rules allow for inelastic resonant scattering of \lya photons, in which the final hyperfine state differs from the initial one \citep{Wouthuysen1952,Field1958}.  This process acts to bring the spin temperature towards the color temperature $T_C$ of the \lya line.  In most situations, $T_C$ is strongly coupled to $T_m$, so the WF effect strengthens the coupling between the matter and spin temperatures.  However, as we will show later, $T_C$ diverges somewhat from $T_m$ at very high redshifts, which alters the predicted value of $T_S$.

Under these three competing effects, the spin temperature is
\be
T_S^{-1}=\frac{T_r^{-1}+x_cT_m^{-1}+x_\alpha T_C^{-1}}{1+x_c+x_\alpha},
\label{SpinTemp}
\ee
where $x_c$ and $x_\alpha$ are the dimensionless coupling strengths of the collisional and WF effects, respectively. The coefficient for collisions with species $i$ is given by
\be
x_c^i=\frac{n_i\kappa_{10}^i}{A_{10}}\frac{T_\star}{T_r},
\ee
where $\kappa_{10}^i$ is the rate coefficient for species $i$, which depends on the kinetic temperature $T_m$. We consider two types of collisions: those between pairs of hydrogen atoms and those between hydrogen atoms and electrons.  Values for the rate coefficients for these two cases can be found in Tables 3 and 4 of Ref.~\cite{Furlanetto2006}, using calculations from Refs.~\cite{Zygelman2005, Sigurdson2006, Furlanetto2007}.

\subsection{The Wouthuysen-Field effect}

The WF coupling coefficient can be written as
\be
x_\alpha=\frac{4P_\alpha}{27 A_{10}}\frac{T_\star}{T_r},
\ee
where 
\be
P_\alpha \equiv \frac{8\pi}{c^2}\int f_\nu\sigma_{\LYA}\nu^2d\nu
\ee
is the total \lya scattering rate per atom, $\sigma_{\LYA}=(\pi e^2/m_e c)f_\alpha\phi_\alpha(\nu)$ is the absorption cross section for the \lya line with oscillator strength $f_\alpha=0.4162$, $\phi_\alpha(\nu)$ is the line profile, and $f_{\nu}$ is the dimensionless photon occupation number.

\begin{figure}
\centering
\includegraphics[width=\columnwidth]{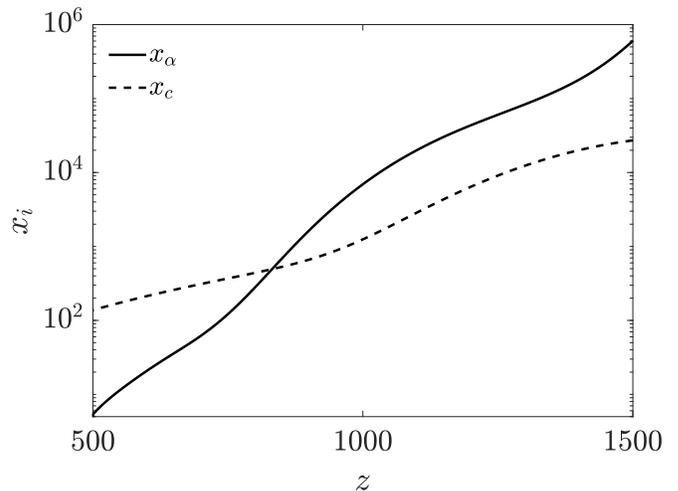}
\caption{Coupling strength as a function of redshift for the WF effect $x_\alpha$ (solid) and collisions between hydrogen atoms $x_c$(dashed).}
\label{fig:xcxa}
\end{figure}

In Figure \ref{fig:xcxa}, we plot the values of the coupling coefficients $x_c$ and $x_\alpha$ as a function of redshift.  As noted by FL13, we find that the WF coupling dominates at $z\gtrsim800$, which motivates an accurate computation of the \lya color temperature at these redshifts. The values of $f_\nu$ used to determine $x_\alpha$ in this plot were computed using the \lya radiative transfer code described below in Section \ref{sec:radtransfer}.

The color temperature is by definition the steady-state spin temperature if the WF field effect was the only spin-changing process, i.e. \cite{Hirata2006}
\be
\left(\frac{n_{1}}{n_{0}}\right)_{\rm ss} \equiv \frac{\Gamma_{01}}{\Gamma_{10}} \equiv 3\rme^{-T_*/T_C} \approx 3 \left(1 - \frac{T_*}{T_C} \right),
\eeq
implying
\be
\frac{T_*}{T_C} = \frac{\Gamma_{10} - \frac13 \Gamma_{01}}{\Gamma_{10}},
\eeq
where $\Gamma_{01}$ and $\Gamma_{10}$ are the rates of \lya scattering events per hydrogen atom changing the hyperfine state $F = 0 \rightarrow 1$, and $1 \rightarrow 0$, respectively. Explicitly, this ratio is 
\be
\frac{T_*}{T_C} = \frac{\int  f_\nu \left[\phi_{10}(\nu) - \phi_{01}(\nu)/3\right]d\nu}{\int f_\nu \phi_{10}(\nu)d\nu}
\ee
where $\phi_{10}$ and $\phi_{01}$ are the relevant thermally-averaged line profiles, which are computed explicitly in Appendix B of \citet{Hirata2006}. They are related through $\phi_{01}(\nu) = 3 \phi_{10}(\nu - \nu_{21})$, so that
\be
\frac{T_*}{T_C} = \frac{\int f_\nu \left[\phi_{10}(\nu) - \phi_{10}(\nu - \nu_{21})\right] d\nu}{\int f_\nu \phi_{10}(\nu) d\nu}.
\ee
Because $\nu_{21}$ is small compared to the Doppler width of the \lya line, we can make the approximation $\phi_{10}(\nu-\nu_{21})\approx\phi_{10}(\nu) - \nu_{21} d \phi_{10}/d\nu$. We then have
\be
T_C\approx\frac{T_\star}{\nu_{21}}\frac{\int f_\nu\phi_{10}(\nu)d\nu}{\int f_\nu \frac{d\phi_{10} }{d\nu} d\nu}=-h\frac{\int f_\nu\phi_{10}(\nu)d\nu}{\int \frac{df_\nu}{d\nu} \phi_{10}(\nu)d\nu},
\label{fullTc}
\ee
where the second equality is obtained by integrating by parts. 

Let us point out that if we make the further approximation that the line is infinitely narrow, $\phi_{10}(\nu)=\delta_D(\nu-\nu_{\LYA})$, this reduces to
\be
T_C\approx-h\left(\frac{d\ln f_\nu}{d\nu}\right)^{-1},
\label{linecenter}
\ee
which is the common expression for the color temperature used in many computations, including those of FL13. 
Except when explicitly stated, we will \emph{not} use Eq.~\eqref{linecenter} to compute the color temperature in what follows, but rather the more accurate Eq.~\eqref{fullTc}.

\begin{figure}
\centering
\includegraphics[width=\columnwidth]{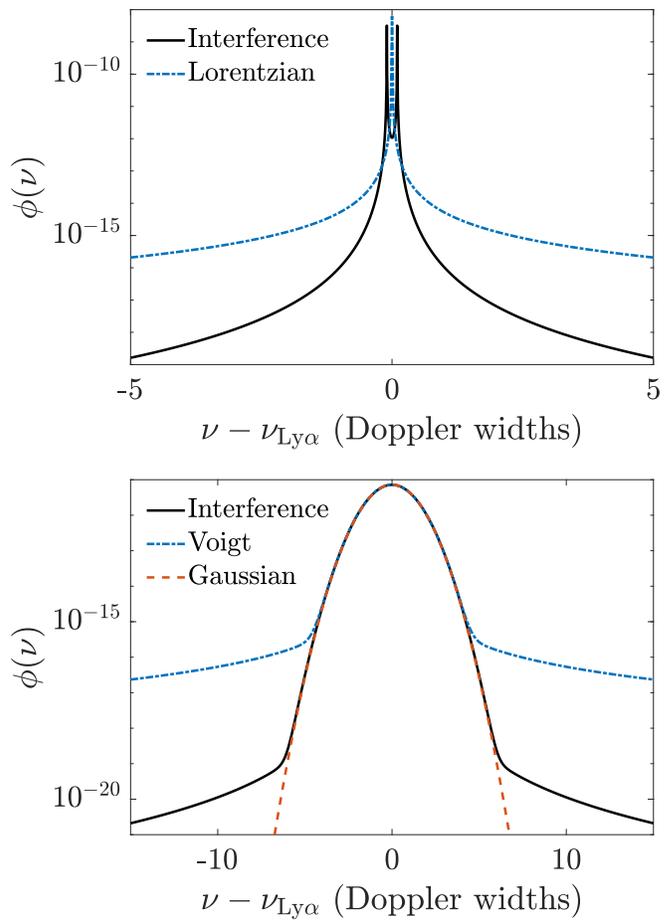}
\caption{\lya line profiles, both unconvolved (top) and convolved (bottom) with the thermal Gaussian profile.  Shown are both the usual Lorentzian/Voigt profile (blue dot-dashed) and the WF interference profile (black solid), along with the Doppler profile (red dashed). All profiles are computed at $z=1000$ using the matter temperature $T_m$ at that redshift.}
\label{fig:profiles}
\end{figure}

Let us now turn our attention to the line profile. A spin-flipping scattering event can proceed through either the $2\rm{p}_{1/2}(F=1)$ or $2\rm{p}_{3/2}(F=1)$ intermediate levels. Far from the \lya resonance, the amplitudes for these two transitions interfere destructively, causing the profiles to differ significantly from the Voigt profile $\phi_\alpha$ which characterizes the overall \lya transition \citep{Hirata2006}. Exact forms for these profiles can be found in Equation (B18) of \citet{Hirata2006}.  The unconvolved line profile is explicitly given by
\be
\phi_{10}=\frac{\gamma\nu_{\rm{FS}}^2}{2\pi\left[(\Delta\nu-\frac{1}{2}\nu_{\rm{FS}})^2+\gamma^2\right] \left[(\Delta\nu+\frac{1}{2}\nu_{\rm{FS}})^2+\gamma^2\right]},
\ee
where $\gamma=50$ MHz is the natural width of the \lya transition and $\nu_{\rm{FS}}=10.9$ GHz is the fine-structure frequency separation.  Compared to the Voigt profile, this interference profile falls off substantially faster in the wings of the \lya line, as $\Delta \nu^{-4}$ instead of $\Delta \nu^{-2}$. This behavior is illustrated in Figure \ref{fig:profiles}, both for unconvolved and convolved profiles.  The two separate fine-structure lines are visible in the unconvolved interference profile, but as $\nu_{\rm{FS}}$ is smaller than a doppler width, the convolved profile is single-peaked.  Just outside the doppler core, the amplitude of interference profile is $\sim4$ orders of magnitude lower than that of the Voigt profile, dramatically reducing the importance of the damping wings.  In fact, we find that using a Gaussian profile alone instead of the interference profile to compute the color temperature obtains the correct result to within $\sim$a few percent.

\section{Global 21cm Signal}\label{sec:globalsignal}

In this section, we will compute the globally averaged 21 cm signal at $z>500$.  Understanding the spin temperature and ultimately the brightness temperature at high redshifts requires accurate computations of several quantities. We obtain the matter temperature and level populations from \textsc{HyRec}. This code does solve for the \lya radiation field, but with a frequency resolution of a few Doppler widths, and only in the damping wings. While this is sufficient for high-accuracy recombination calculations, it may not be enough for our purposes, so we write our own supplemental high-resolution radiative transfer code to handle the region around $\nu_{\LYA}$.  We will also use the free-election fraction from \textsc{HyRec} to compute the effect of free-free processes on our results and produce a complete estimate of observed brightness temperature contrast at a given observing frequency.

\subsection{Lyman-$\alpha$ Radiative Transfer} \label{sec:radtransfer}

To compute the color temperature, we need a high-precision prediction for the spectrum $f_\nu$ in the \lya region.  The radiative transfer equation we need to solve for this spectrum is \citep{Hirata2008, Hirata2009, AliHaimoud2010}
\be
\frac{\partial f_\nu}{\partial t} = H\nu\frac{\partial f_\nu}{\partial \nu} +\dot{f_\nu}|_{\rm{em}}+\dot{f_\nu}|_{\rm{ab}}+\dot{f_\nu}|_{\rm{sc}}.
\ee
In the order written above, the evolution of the spectrum is determined by Hubble drift, true emissions and absorptions of \lya photons, and scatterings.  Absorptions and emissions change the total number of photons, while scatterings and the Hubble drift conserve the photon number but shift their frequencies.

We evolve the spectrum using the method of Ref.~\cite{Hirata2009}, hereafter HF09.  We briefly summarize the computation here, with further details presented for the convenience of the reader in Appendix \ref{app:numerical}.  We divide the frequency range around the Ly$\alpha$ line into bins of constant width $\Delta\ln\nu$.  In a bin centered at frequency $\nu_i$ we evolve the quantity 
\be
N_i=\frac{8\pi\nu_i^3\Delta\ln\nu}{c^3n_H}f(\nu_i),
\label{Ntof}
\ee
which gives the number of photons per hydrogen atom in a given bin.  We use $10^5+1$ bins centered at $\nu_{\LYA}$ with widths $\Delta\ln\nu=10^{-7}$.  These bins are considerably smaller than those used by HF09, but this fine resolution is necessary to produce a converged final spectrum (for a discussion of convergence tests, see Appendix \ref{app:convergence}).  We use a logarithmic step in scale factor equal to our logarithmic frequency interval, $\Delta \ln a = \Delta \ln \nu$. This allows for a simple implementation of Hubble drift: at each timestep we simply redshift the photon numbers by one frequency bin.  After each step, we fill in the boundary bin on the blue side from the output of \textsc{HyRec}.  We start our radiative transfer with a blackbody initial condition at $z=1600$ with temperature equal to the radiation temperature.

The true emission rate is computed based on the population of the various excited states of a hydrogen atoms along with their decay rates.  Level populations are obtained from \textsc{HyRec}, and a correction factor is applied to take into account various photon phase space factors and other non-resonant effects. The absorption rate is obtained from the emission rate by detailed balance, assuming the low-lying $ns$ and $nd$ states are in Boltzmann equilibrium with $2p$ at temperature $T_r$.

The spectrum of photons near the \lya transition is smooth on the scales of a doppler width (see e.g.~the discussion in \cite{AliHaimoud2010}), so we can evolve the spectrum using the Fokker-Planck diffusion approximation \cite{Rybicki1994}. In principle, we should consider all six of the different transitions described in Ref.~\cite{Hirata2006} for different hyperfine levels. However, as stated above, all of these different lines fall well within the \lya Doppler width at $z\gtrsim 500$, so we will again treat the \lya line as a single transition with a single Voigt profile (the interference term in $\phi_{10}$ does not appear in the full \lya profile). We only account for kinetic diffusion, and neglect the spin diffusion term considered in Ref.~\cite{Hirata2006}. This is justified because the hyperfine splitting is much smaller than a Doppler width at the redshifts of interest. We use a tri-diagonal discretized diffusion operator, for which we enforce detailed balance at the matter temperature $T_m$. This translates the fact that \lya photons mostly exchange energy with the kinetic degrees of freedom of the gas during scattering events (we will get back to the exchange of energy with the spin degrees of freedom below). We evolve the photon field at each timestep using an implicit Euler method, after which we shift each value redward to account for Hubble redshifting.

\begin{figure}
\centering
\includegraphics[width=\columnwidth]{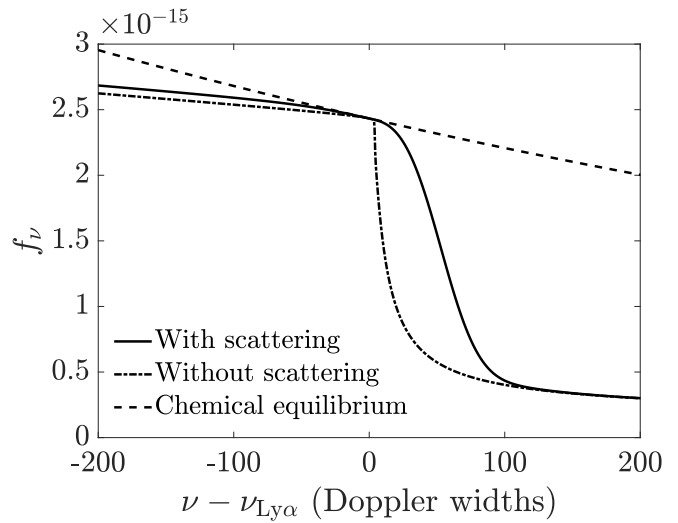}
\caption{Spectra near the \lya transition at $z=1000$, calculated assuming the full diffusion behavior (solid), neglecting scattering effects (dot-dashed), and assuming equilibrium (dashed).  Values on the x-axis give the number of doppler widths from line center.}
\label{fig:spectrum}
\end{figure}

The result of our computation at $z=1000$ is plotted in Figure \ref{fig:spectrum}.  The solid curve shows the spectrum including all of the effects discussed above, while the dot-dashed curve shows the effect of neglecting scattering.  As expected, both spectra approach the equilibrium spectrum $f_\nu^{\rm{eq}}\approx(x_{2p}/3x_{1s})\exp\left[-h(\nu-\nu_{\LYA})/T_r\right]$ (shown by the dashed curve) within a few doppler widths of the line center.  Note that, though the difference cannot be seen by eye, scattering actually drives the spectrum towards equilibrium with the kinetic degrees of freedom of matter at temperature $T_m$ rather than $T_r$. See Figure \ref{fig:bigF} for more details.

\subsection{\lya Color temperature}
\begin{figure}
\centering
\includegraphics[width=\columnwidth]{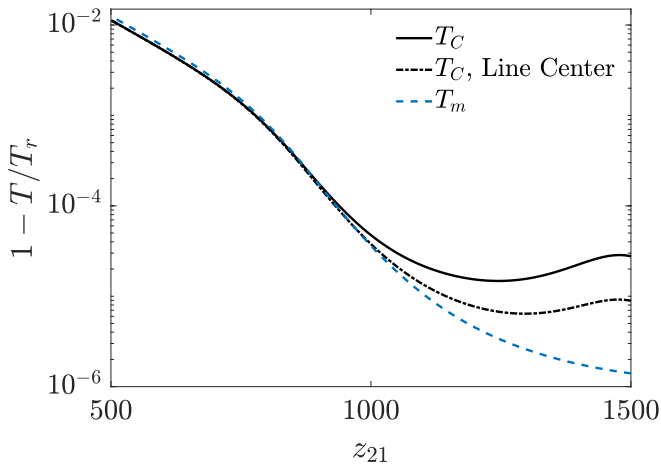}
\caption{Deviation of color temperature $T_C$ from radiation temperature $T_r$ as a function of redshift, computed both using the full \lya spectrum as in Eq.~\eqref{fullTc} (black solid) and only the line center as in Eq.~\eqref{linecenter} (black dot-dashed).  Shown for comparison (blue dashed) is the deviation between the matter temperature $T_m$ and the radiation temperature $T_r$.}
\label{fig:Tcolor}
\end{figure}

Now that we have our \lya spectrum, we can use the results from Section \ref{sec:background} to compute the color temperature.  In Figure \ref{fig:Tcolor} we plot the deviation of the color temperature from the radiation temperature as a function of redshift.  The blue solid curve shows the result evaluated by applying Equation \ref{fullTc} to the spectrum computed in the previous section.  We have also plotted for comparison the matter temperature output from \textsc{HyRec} (red dashed curve).  From this it can clearly be seen that at $z\sim500$, the usual assumption that $T_C\sim T_m$ is quite accurate.  However, at $z\gtrsim 1000$, the color temperature difference approaches a plateau, differing from $T_r$ by a few parts in $10^5$.  At the highest redshifts we consider this effect has become quite significant, as the color temperature contrast is about an order of magnitude higher than that of the matter temperature.

We also plot for the sake of comparison the color temperature obtained using the narrow-line approximation from Equation \ref{linecenter} (blue dot-dashed curve).  The color temperature calculated in this limit tracks the matter temperature to much higher redshifts, plateauing to a value a few times lower than that of the full computation.  This suggests that photons outside of the very center of the line contribute significantly to the deviation of the color temperature from the gas temperature.

We can confirm this conclusion by looking closer at our computed spectrum.  We start by using Eq.~(\ref{fullTc}) to write
\be
1-\frac{T_C}{T_m} \approx \frac{T_m}{T_C} - 1 \approx \int \mathcal{F}(\nu)d\nu
\label{TcF}
\ee
where we have defined
\be
\mathcal{F}(\nu)\equiv -  \frac1{f_{\nu_{\alpha}}}\left[f_\nu + \frac{T_m}{h} \frac{df_\nu}{d\nu}\right] \phi_{10}(\nu) .
\label{Fdef}
\ee
It can easily be seen that $\mathcal{F}(\nu)=0$ and $T_C=T_m$ if the spectrum is a blackbody at temperature $T_m$ (in the Wien-tail approximation).

\begin{figure}
\centering
\includegraphics[width=\columnwidth]{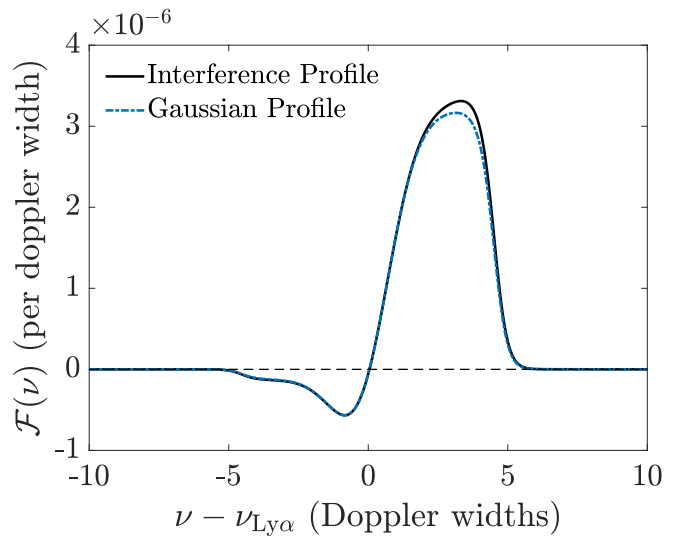}
\caption{The relative contribution from different frequencies to the difference between $T_C$ and $T_m$ at $z=1000$, illustrated using the quantity $\mathcal{F}(\nu)$ defined in Equation (\ref{Fdef}).  The black solid curve shows the result using the full convolved interference profile, while the blue dot-dashed curve shows the result obtained from a Gaussian profile alone.  Both curves are normalized such that their integral yields the correct value of $1-T_C/T_m$.}
\label{fig:bigF}
\end{figure}

Figure \ref{fig:bigF} uses the quantity $\mathcal{F}(\nu)$ to illustrate the source in frequency space of the deviation between color and matter temperatures. The blue curve shows $\mathcal{F}(\nu)$ (multiplied by a Doppler width) for the same spectrum used in Figure \ref{fig:Tcolor} at $z=1000$, roughly the point where $T_C$ begins to diverge from $T_m$. Results are shown using both the full interference profile and the doppler profile alone.  A sizable excess can be seen in the five or so doppler widths blueward of the line center, translating to the excess over the line center value seen in Figure \ref{fig:Tcolor}.  Note that the values of $\mathcal{F}$ for both profiles are quite similar.  They yield the same value of $1-T_C/T_r$ to within $3\%$, justifying our use of a Gaussian profile in our full calculation. 

\subsection{Cosmological free-free effects}
\label{sec:freefree}

At the redshifts we consider here, the large free-electron abundance means that we must consider the effects of free-free processes on our spectrum.  We will quantify the strength of these effects here. 

We denote by $\alpha_{\rm ff}(\nu,z)$ the net free-free absorption coefficient (including stimulated emission) for photons with frequency $\nu$ at redshift $z$.  For thermal Bremsstrahlung where the electron temperature is equal to the matter temperature, we have \citep{Rybicki1986}
\be
\alpha_{\rm ff}(\nu,z) = \frac{4e^6}{3m_ec\nu^{2}}\left(\frac{2\pi}{3m_eT_m^3}\right)^{1/2} n_e^2~ g_{\rm ff}(\nu, T_m),
\ee
where $n_e$ is the number density of free electrons (or protons, assuming that Helium is fully recombined). We obtain values for the free-free Gaunt factor $g_{\rm ff}$ from \cite{vanHoof2014}.  Note that, as in our previous calculations, we are working in the Rayleigh-Jeans regime.

We denote by $T_b(\nu_0, z)$ the brightness temperature at frequency $(1+z)\nu_0$ and redshift $z$, and $t_b(\nu_0, z) \equiv [T_b(\nu_0, z)-T_r(z)]/(1+z)$. We also define $t_m(z) \equiv [T_m(z)-T_r(z)]/(1+z)$ and $t_s(z) \equiv [T_s(z)-T_r(z)]/(1+z)$. In the Rayleigh-Jeans regime of interest, the radiative transfer equation can be rewritten as the following differential equation for $t_b$, at constant $\nu_0$:
\be
\frac{d t_b}{dz} = \frac{d \tau_{\rm ff}}{dz}(t_b - t_m) + \frac{d \tau_{21}}{dz} (t_b - t_s),
\eeq
where the differential free-free optical depth is
\be
\frac{d \tau_{\rm ff}}{dz}(\nu_0, z) \equiv \frac{\alpha_{\rm ff}(\nu_0(1+z), z)}{(1+z) H(z)},
\eeq
and, approximating the 21-cm line as narrow, 
\be
\frac{d \tau_{21}}{dz}(\nu_0, z) \equiv \tau_{21}(z_{21}) \delta_{\rm D}(z - z_{21}), \ \ \ \ 1+z_{21} \equiv \frac{\nu_{21}}{\nu_0}.
\eeq
In this equation, $z_{21}$ is the redshift at which a photon of frequency $\nu_0$ today was resonant with the 21-cm hyperfine transition.

This equation can be solved in three steps:\\
$(i)$ Assuming $t_b(z = \infty) = 0$, the brightness temperature just prior to resonance is
\be
t_b(z_{21}^+) = t_{\rm ff} (\nu_0, z_{21}),
\eeq
where
\barr
t_{\rm ff} (\nu_0, z) &\equiv& \int_z^{\infty} dz'~\exp\left[-\tau_{\rm ff}(\nu_0;  z, z')\right] \frac{d \tau_{\rm ff}}{dz'} t_m(z'), ~~~~~\\
\tau_{\rm ff}(\nu_0;  z, z') &\equiv& \int_{z}^{z'} dz'' \frac{d \tau_{\rm ff}}{dz''}(\nu_0, z'').
\earr
$(ii)$ Assuming $\tau_{21} \ll 1$, the change in brightness temperature after resonant interaction with the 21-cm line is
\be
t_b(z_{21}^-) \approx t_b(z_{21}^+)(1-\tau_{21}) + \tau_{21} t_s \approx t_b(z_{21}^+) + \tau_{21} t_s.
\eeq
$(iii)$ After resonant interaction with the 21-cm line, the radiation changes through free-free processes alone:
\barr
T_{21}(\nu_0) &=& t_b(\nu_0, 0) = t_b(z_{21}^-) \exp\left[- \tau_{\rm ff}(\nu_0; 0, z_{21})\right] \nonumber\\
&&+ \int_0^{z_{21}} dz ~\exp\left[-\tau_{\rm ff}(\nu_0;  0, z)\right] \frac{d \tau_{\rm ff}}{dz} t_m(z).
\label{tau_post21}
\earr
Sewing everything together, we arrive at the following brightness temperature contrast at redshift zero,
\barr
T_{21}(\nu_0) = \tau_{21} t_s(z_{21})\exp\left[- \tau_{\rm ff}(z_{21})\right] + t_{\rm ff} (\nu_0, 0),~~~~
\label{fullT21}
\earr
where, for short, we have defined
\barr
\tau_{\rm ff}(z_{21}) &\equiv& \tau_{\rm ff}(\nu_0; 0, z_{21}).
\earr
So the effect of free-free absorption on the global signal is twofold: $(i)$ it exponentially damps the standard contribution by $\rme^{-\tau_{\rm ff}}$, and $(ii)$, it adds a frequency-dependent offset $t_{\rm ff}(\nu_0, 0)$. We can rewrite the latter as the following weighted integral of $t_m$:
\be
t_{\rm ff} (\nu_0, 0) \equiv \int_0^{\infty} dz~V_{\rm ff}(\nu_0, z) t_m(z), \label{eq:tff}
\ee
where 
\be
V_{\rm ff}(\nu_0, z) \equiv \exp\left[- \tau_{\rm ff}(\nu_0; 0, z)\right] \frac{d \tau_{\rm ff}}{d z}(\nu_0, z).\label{ffVisibility}
\ee
The function $V_{\rm ff}$ plays the role of a visibility function similar to that of CMB anisotropies: at fixed $\nu_0$ its integral over redshift is unity, and it weighs the contribution of the source function $t_m(z)$ in Eq.~\eqref{eq:tff}.

\begin{figure}
\centering
\includegraphics[width=\columnwidth]{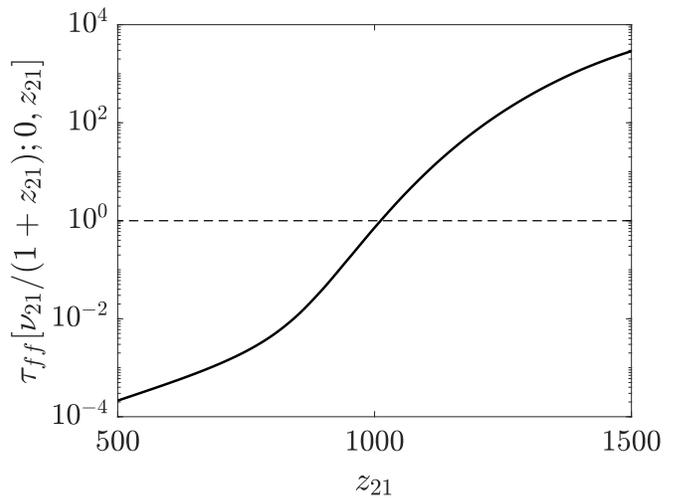}
\caption{Free-free optical depth $\tau_{\rm{ff}}$ between redshifts 0 and $z_{21}$ for photons with frequency $\nu_{21}$ at $z_{21}$.  The free-free process becomes optically thick (marked with dashed line) at roughly $z\sim1000$.}
\label{fig:tauFF}
\end{figure}

Figure \ref{fig:tauFF} shows the value of $\tau_{\rm{ff}}$ between redshifts 0 and $z_{21}$ for photons emitted at $\nu_0=\nu_{21}/(1+z_{21})$, computed using the output of \textsc{HyRec}. For photons emitted well after recombination, the optical depth is small because the ionization fraction is small. The optical depth then rises rapidly following the rising ionization fraction, reaching unity around $z\sim1000$ and several thousand by $z\sim1500$.  

\begin{figure}
\centering
\includegraphics[width=\columnwidth]{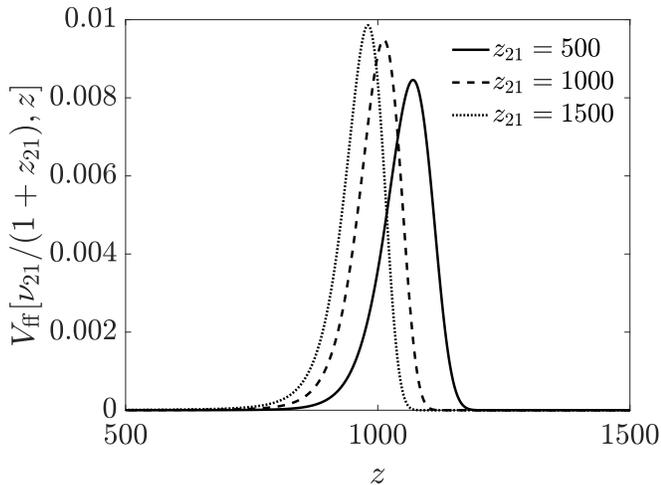}
\caption{Free-free visibility functions $V_{\rm ff}$ for instruments targeting redshifts $z_{21}$=500 (solid), 1000 (dashed), and 1500 (dotted).}
\label{fig:VFF}
\end{figure}

Figure \ref{fig:VFF} shows the visibility function $V_{\rm ff}$ defined in Eq. \eqref{ffVisibility} for instruments targeting $z_{21}=500$, 1000, and 1500.  One can easily show [e.g.~for a power-law $\tau_{\rm ff}(z)$] that the visibility function peaks near $\tau_{\rm ff} \sim 1$, i.e~a surface of last free-free interaction, by analogy to the CMB surface of last scattering.  As expected, our plotted $V_{\rm ff}(z)$'s all peak near $z\sim1000$, roughly where we expect $\tau_{\rm ff}$ to be unity.  The peaks shift slightly because each curve deals with photons at different frequencies.  As one increases the redshift $z_{21}$ of resonance (hence decreasing $\nu_0$), the free-free optical depth increases, and the free-free visibility function peaks at lower redshifts, where $|t_m|$ is larger.  We therefore expect $|t_{\rm ff}(\nu_0,0)|$ to increase with $z_{21}$.  In other words, the impact of free-free effects on our signal should be higher for instruments targeting higher redshifts.

\subsection{Galactic free-free effects}

The plasma in the Milky Way ISM is also a source of free-free opacity at low frequencies, which can obscure the cosmological signal. The free-free opacity can be modeled as
\begin{equation}
\tau_{\rm ff,Gal}(\nu) = \left( \frac{\nu_\star}{\nu} \right)^{2.1},
\end{equation}
where $\nu_\star$ is a transition frequency where the optical depth is unity, which depends on the line of sight (lines of sight with longer path length, or through denser or colder gas, will have more absorption and a larger $\nu_\star$). The power law deviates slightly from 2 due to the Coulomb logarithmic factor in the Gaunt factor. Modeling of the radio spectrum of the Galactic poles -- which is dominated by synchrotron emission and free-free absorption -- suggests $\nu_\star \approx 2.2$ MHz (e.g., Ref.~\cite{1979MNRAS.189..465C}; this is $F=5$ in their notation). This means that at $z\gtrsim 650$, the Milky Way ISM is optically thick to the cosmological 21 cm signal in the sense that $\tau_{\rm ff,Gal}>1$; however, in the range considered in this paper the optical depth is only a few, and so some signal survives and reaches an observer inside the Milky Way.
Furthermore, it seems likely that the free-free absorption will turn out to be patchy when the MHz sky is observed at high resolution instead of with single antennas, with dense structures corresponding to peaks in $\tau_{\rm ff,Gal}$. (This is seen in the Galactic Plane in the 36--73 MHz band, e.g.\ \cite{2017arXiv171100466E}.) This means that the optical depth inferred from wide-angle temperature measurements may be an underestimate on some lines of sight, and an overestimate on others.

\subsection{Results}
\label{sec:results}

\begin{figure}
\centering
\includegraphics[width=\columnwidth]{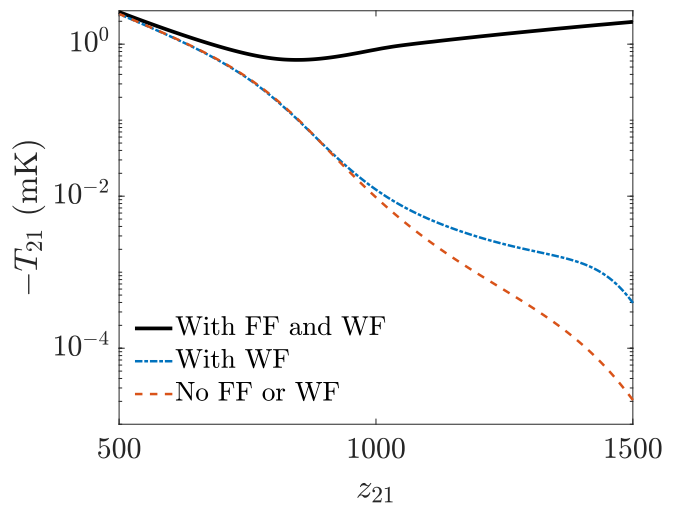}
\caption{Predicted brightness temperature in the 21 cm line as a function of redshift of resonance.  Results are plotted using our full computation (solid black), as well as in the limits where free-free (dot-dashed blue) and both free-free and WF (dashed red) are neglected.  Note that the $y$-axis shows the negative of the true brightness temperature contrast. Note that, though we have plotted all three curves as a function of $z_{21}$, unlike the other two the black curve includes free-free contributions from the entire line of sight, not just a narrow band around $z_{21}$.}
\label{fig:global}
\end{figure}

We plot in Figure \ref{fig:global} the predicted global observed brightness temperature $T_{21}$ as a function of redshift of resonance $z_{21}$.  Note that for ease of viewing we have plotted $-T_{21}$ so that, though $T_b<T_r$, a stronger signal still appears higher on the $y$-axis. The dashed red curve shows the results which would be obtained neglecting both the WF effect (by setting $x_\alpha=0$ in Eq.~\eqref{SpinTemp}) and free-free effects computed above. The blue dot-dashed curve shows the effect of adding in the WF coupling with our computed $T_C(z_{21})$.  As expected based on Figure \ref{fig:Tcolor}, this has the effect of substantially increasing the amplitude of the observed signal beyond $z_{21}\sim 1000$.  The solid black curve shows the full result of evaluating Equation (\ref{fullT21}) including free-free.  At $z_{21}=500$, the 21 cm signal still dominates over the free-free contribution.  However, as predicted in the previous section, the amplitude of the free-free effect increases with redshift, overwhelming the 21 cm component for $z_{21}\gtrsim800$.

It should also be noted that all three curves in Figure \ref{fig:global} converge at $z\sim500$.  Below this redshift, the WF and free-free effects we consider here become subdominant to collisional coupling to the cooling hydrogen atoms.  Therefore, past computations of the 21 cm signal for $z\lesssim500$ should not be significantly affected by the processes we consider here.

In conclusion of this section, we found that the \emph{global} (monopole) distortion to the CMB at frequencies $\nu \lesssim 2$ MHz is mostly determined by free-free interactions, and has little dependence on the 21-cm transition. We now turn on to the study of fluctuations.

\section{21 cm fluctuations}
\label{sec:fluctuations}

We now turn our attention to the fluctuations in the 21 cm signal from the recombination epoch, which could in principle be detectable with interferometers.

\subsection{Accessible scales}

The derivation above still applies to the case where the optical depths and temperature are spatially-dependent functions. Our first consideration is the range of radial ($k_\parallel$) and transverse ($k_\perp$) wavenumbers that are {\em in principle} observable. In this section, we define the re-scaled redshift variable
\begin{equation}
z_3 \equiv \frac{1+z}{1000}
\end{equation}
for simplicity of notation.

\subsubsection{Radial wavenumbers} \label{sec:radial}
In the radial direction, we recall that a sinusoidal fluctuation in brightness temperature consisting of $N$ oscillations per $e$-fold in frequency has a radial wavenumber
\begin{equation}
k_\parallel = 2\pi N\frac{H(z)}{c(1+z)} = 0.024N z_3^{1/2} {\rm Mpc}^{-1}.
\end{equation}
If one fits and removes a spectrally smooth foreground model, then modes with low $N$ are necessarily lost. The minimum $N$ that survives depends on the complexity of the foreground model required and the bandwidth over which it can be fit. Even in the extraordinarily over-optimistic case that over an on octave-wide bandwidth ($\Delta\ln\nu = 0.7$) a quartic polynomial (2 oscillations) would suffice, we would have $N_{\rm min}\approx 2/0.7$ and $k_{\parallel,\rm min} \sim 0.07z_3^{1/2}\,$Mpc$^{-1}$. Realistic foregrounds, and consideration of spectral structure imprinted by the instrument, may set a much larger value of $k_{\parallel,\rm min}$.

\subsubsection{Transverse wavenumbers}
In the transverse direction, scattering in the Milky Way's ISM provides a serious obstacle to observations at low frequencies. This scattering smears any cosmological signal into a seeing disk, or equivalently suppresses the electric field correlation function at long baselines. The NE2001 model \cite{2002astro.ph..7156C} gives a scattering measure of $8.5\times 10^{-5}$ ($1.8\times 10^{-4}$) kpc m$^{-20/3}$ at the North (South) Galactic Pole. This gives an angular broadening of $\theta = 0.0041z_3^{2.2}$ {\rm radians} at the NGP (the coefficient is 0.0064 at the SGP). This implies a cutoff at the scale
\begin{equation}
k_{\perp,\rm max} = \frac{2.355}{\theta D(z)}
= 0.041 z_3^{-2.2} {\rm Mpc}^{-1}
\label{eq:kperpmax}
\end{equation}
at the NGP (the coefficient is 0.026 at the SGP).  Fourier modes with larger transverse wavenumbers than this are washed out. There is quite a bit of uncertanties in these numbers, since the scattering model has not been validated as such low frequencies; nevertheless, the qualitative picture that angular broadening is enormous in the MHz regime seems inescapable.

Regardless of the direction of the Fourier mode, there is a maximum wavenumber -- the baryon Jeans scale $k_{\rm Jeans}$ -- beyond which the cosmological signal decreases rapidly. Given Eq.~\eqref{eq:kperpmax}, this is only relevant in the radial direction, and sets $k_{\parallel,\rm max}\sim k_{\rm Jeans}$.

\subsection{Estimation of the signal strength}

We estimate the characteristic fluctuation $\dtb$ of these fluctuations as the square root of the dimensionless 21-cm power spectrum, using a method similar to that of FL13. To facilitate comparison with their results, we neglect perturbations to the ionization fraction and to the gas temperature. The fluctuation amplitude is then determined by fluctuations in the optical depth, which only depend on the baryon density and velocity perturbations, $\delta_b$ and $\delta_v\equiv(1+z)\ dv_\parallel/dr_\parallel/H(z) -1$, respectively, as well as perturbations in the spin temperature. In Fourier space, and assuming linear perturbations, $\delta_v$ is given by the linearized continuity equation, implying
\be
\delta_v=-\mu^2\kappa\delta_b, \ \ \ \ \mu \equiv \hat{k} \cdot \hat{n} = k_{||}/k,
\ee
where $\hat{n}$ is the line of sight, $\hat{k}$ is the direction of the Fourier wavenumber and $\kappa\equiv |d\ln\delta_b/d\ln(1+z)|$ is the growth rate of baryonic perturbations. Typically, we find $\kappa \sim 5$ for the wavelengths and redshifts of interest. This is larger than unity (as assumed in FL13) because near recombination, baryons start nearly unclustered and quickly catch up to the dark matter. Using Eq.~\eqref{eq:T21-num}, we see that the 21-cm brightness contrast is proportional to $(1+\delta_b-\delta_v)(1 - T_r/T_s) = \left(1 + (1 + \mu^2 \kappa) \delta_b\right) (1 - T_r/T_s)$. 

Because the coupling $x_c$ between $T_m$ and $T_s$ increases with baryon density, we can write $x_c=\overline{x}_c(1+\delta_b)$.  After some algebra, we can then say that
\be
1-T_r/T_s=\overline{1-T_r/T_s}~(1+X\delta_b),
\ee
where
\be
X\equiv\frac{\overline{x}_c(T_m - T_r)}{\overline{x}_c(T_m - T_r) + x_\alpha(T_c - T_r)} - \frac{\overline{x}_c}{1 + \overline{x}_c + x_\alpha},
\ee
and we have made the approximation $1- T_r/T_m \approx T_m/T_r -1$.  The intensity fluctuation amplitude is then
\barr
\dtb &=& \Dtb(1+\delta_b-\delta_v+X\delta_b) \nonumber \\
&=&(1+X+\mu^2\kappa)\delta_b.
\earr
We need now only one more component for our fluctuation amplitude: the effect of Bremsstrahlung. The line-of-sight integral in Eq.~\eqref{eq:tff} implies that the second term in Eq.~\eqref{fullT21} ought to have little fluctuations on scales smaller than the width of the visibility function, corresponding to wavenumbers $k_{||} \gtrsim 0.05$ Mpc$^{-1}$. Given the lower limit on observable radial wavenumbers discussed in Section \ref{sec:radial}, we conclude that foreground cleaning will remove any observable fluctuations of the second term in Eq.~\eqref{fullT21}. Fluctuations in the first term do survive, albeit supressed by free-free opacity:
\be
\dtb^{\rm{obs}}=\dtb^0e^{-\tau_{\rm{ff}}}.
\ee
This will exponentially suppress any fluctuations beyond $z\sim1000$.  

\begin{figure}
\centering
\includegraphics[width=\columnwidth]{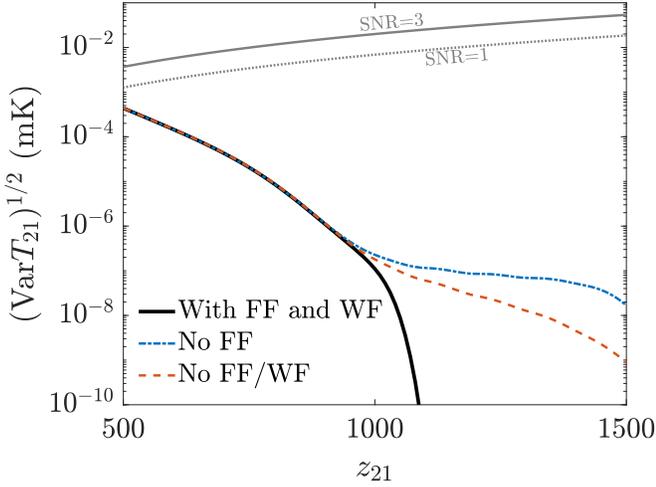}
\caption{Amplitude of the 21-cm brightness fluctuations integrated over the observable range of scales under our full calculation (solid black curve), neglecting free-free effects (blue dot-dashed), and neglecting both free-free and WF effects (red dashed).  Also plotted are the minimum amplitudes detectable with an idealized instrument with $S/N=3$ (grey solid) and $S/N=1$ (grey dotted).}
\label{fig:fluct}
\end{figure}

Figure \ref{fig:fluct} shows the amplitude of the temperature fluctuations integrated over the range of accessible scales defined in the previous section.  Specifically, we plot the square root of the variance of the fluctuations,
\barr
\left({\rm Var}\,T_{21}\right)^{1/2}
\!\!\!\!\!\!\!\!\!\!\!\!\!\!\!\!\!\!\!\!\!\!\!\!
&&
\nonumber \\
&=&\left[\int P_{21}(k)\frac{d^3k}{(2\pi)^3}\right]^{1/2} \nonumber \\
&=&\left[\frac{T_{21}^2}{(2\pi)^2}\int k_{\perp}(1+X+\mu^2\kappa)^2P_bdk_\perp dk_\parallel\right]^{1/2},~~~~~~
\earr
where we have computed the baryon power spectrum $P_b(k)$ using CAMB \citep{Lewis2002}.  In addition to the results of our full computation, we have also plotted amplitudes with the free-free and WF effects removed.  Though the WF effect acts to increase the amplitude over the collision-only prediction, this contribution only becomes significant above $z\sim1000$, in the free-free-suppressed regime.  Thus we see that 21-cm brightness temperature fluctuations cannot be observed even in principle beyond recombination, despite the fact that there is a non-vanishing (albeit small) global signal.

\subsection{Idealized signal-to-noise ratio}

We may now consider whether the recombination-era signal is detectable by a plausible future experiment. Due to the frequency range, this would have to be in space, either free-floating or possibly on the Moon.

For an ideal densely filled array of collecting area $A$, observing a region of solid angle $\Delta\Omega$ large compared to the beam size ($\lambda_{\rm obs}^2/A$), and averaging over a range of frequencies $\Delta\nu$ for an observation time $t_{\rm obs}$, the idealized uncertainty in the brightness temperature is
\begin{equation}
\sigma_T = \sqrt{\frac2{N_{\rm modes}}}\, T_{\rm sky}
= \frac{\lambda_{\rm obs}}{\sqrt{2A\,\Delta\Omega \,\Delta\nu\,t_{\rm obs}}}\, T_{\rm sky},
\end{equation}
where $T_{\rm sky}$ is the sky temperature, $\lambda_{\rm obs}$ is the observed wavelength, and $N_{\rm modes}$ is the total number of modes of the electromagnetic field collected by the experiment within that range of frequencies and observing directions (and counting both polarizations). Based on the {\slshape Radio Astronomy Explorer 2} maps \cite{1978ApJ...221..114N}, we obtain a sky temperature near the Galactic poles of 8 MK (2.2 MHz) or 13 MK (1.31 MHz), which treated as a power law would give a brightness temperature of $T_{\rm sky} = 12z_3^{0.94}$\,MK. The observed wavelength is $\lambda_{\rm obs} = 0.21z_3\,$km. Then:
\begin{equation}
\sigma_T = 6.9\times 10^{-3}\left( \frac{A}{A_{\rm L}} \,
\frac{\Delta\Omega}{\pi\,\rm sr} \,
\frac{\Delta\nu/\nu}{0.5} \frac{t_{\rm obs}}{100\,\rm yr} \right)^{-1/2} z_3^{2.44}\,{\rm mK},
\label{eq:sigmaT}
\end{equation}
where we have scaled the area relative to $A_{\rm L} = \pi R^2_{\rm L} = 9.5\times 10^6\,{\rm km}{^2}$, the cross-sectional area of the Moon.

Let us now consider a signal on the sky with variance ${\rm Var}\, T_{21}$. Suppose that we divide a survey into $N_{\rm 3D}$ 3D pixels in frequency-angular position space of volume $\Delta\nu_1\,\Delta\Omega_1$ (units: Hz sr), each with a signal $T_{21}(\nu,\Omega)$. To obtain the most optimistic $S/N$ calculation, we suppose that (somehow) the relevant modes of the cosmological density field were already measured and we attempted a 21 cm detection by cross-correlation against a perfect template. (Given that the only other tracer at these redshifts is the CMB itself, which provides only 2D information, such a perfect template may not exist; we therefore view this calculation as a way to obtain an optimistic bound to the $S/N$ ratio.) In this case, the total $S/N$ ratio would be
\begin{eqnarray}
\left(\frac SN\right)^2 &=& \sum_{\rm 3D\,pix} \left[\frac{T_{21}(\nu,\Omega)}{\sigma_T(\Delta\nu_1,\Delta\Omega_1)}\right]^2
\nonumber \\
&=& N_{3D}\frac{{\rm Var}\,T_{21}}{\sigma_T^2(\Delta\nu_1,\Delta\Omega_1)}
\nonumber \\
&=& \frac{{\rm Var}\,T_{21}}{\sigma_T^2(\Delta\nu_{\rm tot},\Delta\Omega_{\rm tot})},
\end{eqnarray}
where the temperature uncertainty in the last line $\sigma^2_T$ is computed according to Eq.~(\ref{eq:sigmaT}) using the bandwidth $\Delta\nu_{\rm tot}$ and the solid angle $\Delta\Omega_{\rm tot}$ of the whole survey volume instead of an individual 3D pixel.

We can see that for a ``full survey'' that consists of one Galactic cap ($\pi$ sr), and taking a wide bin width in redshift or frequency $\Delta\nu/\nu=\Delta z/(1+z) = 0.5$, assuming the existence of a perfect cross-correlation template, assuming that we can densely fill the entire Moon with receivers and acquire 100 years of good data, and assuming perfect foreground subtraction, a signal with $({\rm Var}\,T_{21})^{1/2} = 0.0069z_3^{2.44}\,$mK could be measured with $S/N=1$ and a signal at $0.02z_3^{2.44}$ mK could be measured with $S/N=3$. 
As shown in Figure~\ref{fig:fluct}, even this very optimistic scenario is not sufficient for a detection of the recombination-era 21 cm signal.


\section{Discussion}
\label{sec:discussion}

Interactions between hydrogen atoms and \lya photons in the very high-redshift Universe create a deviation between the spin temperature and the CMB brightness temperature, leading to a small but in principle measurable 21-cm signal at $z\gtrsim500$.  When we take into account an accurate estimate of the WF effect, we find that it along with coupling with CMB photons and collisions between hydrogen atoms produce a global 21-cm amplitude of roughly $\sim2.5\ \rm{mK}$ at $z\sim500$, falling to $\sim 0.01\ \rm{mK}$ at $z\sim1000$.  Free-free effects along the line of sight to the target redshift then increase the amplitude of the global brightness temperature deviation to $\sim1-2\ \rm{mK}$ at $z\sim1000$ and beyond as photons redshift through successive shells of cooling free electrons.

This may make it appear at first glance that there is an observable 21-cm global signal beyond recombination.  Unfortunately, once a continuum process such as Bremsstrahlung begins to dominate, 
it is no longer possible to isolate emission from a single redshift slice, as the Bremsstrahlung contribution comes from integrating conditions along the entire line of sight.  

However, this same continuum behavior which makes it impossible to obtain global tomography at these high redshifts means that the contribution from Bremsstrahlung to 21 cm fluctuations can be subtracted along with usual 21 cm foregrounds.   At $z\sim1000$ we predict small and likely unobservable perturbations with an amplitude of $\sim10^{-7}$ mK. Even this small signal is quickly drowned out by free-free absorption beyond $z\sim1100$, as the ionization fraction and as a consequence the free-free optical depth steeply increase.  This therefore is the ultimate redshift limit for 21 cm cosmology, beyond which no signal can be recovered.

Previous work on this topic by FL13 obtained very different results for both the global 21-cm amplitude and the size of the fluctuations.  They predict a $\sim1\ \rm{mK}$ global signal at $z\sim1000$ from the WF effect alone, along with a fluctuation amplitude of $\sim10^{-4}\ \rm{mK}$. These are $\sim100$ times larger than our estimates in this redshift range. These differences arise primarily from our more accurate treatment of the Ly$\alpha$ radiative transfer, hence of the Ly$\alpha$ color temperature. Indeed, we implemented a high-resolution radiative transfer code, producing a considerably more accurate prediction of the spectrum near the \lya transition than is possible using \textsc{HyRec} alone. In addition, we use the full frequency range of the \lya profile when computing the color temperature, rather than just the slope at line center, which underestimates the brightness temperature contrast at high redshifts. Finally, our fluctuation amplitude is slightly enhanced because we use the numerically-computed growth rate for the baryon perturbations. 

There are still several effects which we have not included in our calculation, though we do not expect any of them to alter our results significantly.  In our radiative transfer calculation, we treated the line profile around \lya as a single Voigt profile, rather than using the full form for the profile computed in \cite{Hirata2006} including all of the different hyperfine transitions.  This should in principle alter both our computation of the \lya spectrum and that of the color temperature.  In addition, as we mentioned above, the WF coupling induces an additional diffusion term into the Fokker-Planck equation which we neglect.  This spin-diffusion term should act to bring the color temperature into equilibrium with the spin temperature.  Including this effect accurately would require recursively solving for the spin and color temperatures together.  However, since we are working at high redshift, the \lya line is fairly wide and both of these effects should be weaker.  Comparison between the results of Refs.~\cite{Hirata2006} and \cite{Chen2004} suggests that both of these corrections should come in a the sub-percent level at our redshifts of interest.  We also neglected the impact of ionization fluctuations on our perturbation amplitudes, which would alter the fluctuation size somewhat.

Though our computations are limited to redshifts above 500, our results have an interesting implication for current measurements of lower redshift 21 cm signals.  In particular, the Experiment to Detect the Global Epoch of Reionization Signature (EDGES) experiment recently reported a deeper-than-expected 21 cm absorption trough at a frequency corresponding to the expected redshift of cosmic dawn \citep{Bowman2018}.  In standard models of 21 cm evolution, it is assumed that \lya photons from the first stars would couple $T_s$ to the adiabatically-decreasing matter temperature at this redshift, but the spin temperature inferred from the EDGES result is significantly lower than the minimum plausible matter temperature. \citet{Barkana2018} proposes that a spin temperature below the matter temperature is evidence for baryon-dark matter interactions cooling the matter faster than expected \cite{Tashiro2014, Munoz2015b}. However, in our work, albeit at very different redshifts, we find a spin temperature outside the range bounded by $T_m$ and $T_r$ driven only by \lya scatterings. Though our signal is quite small, it motivates further investigation of other subtle radiative transfer effects to determine whether unexpected results could be produced through more conventional means.


\section{Conclusion}
\label{sec:conclusion}

In the wake of the first possible detection of a 21-cm signal at $z\sim 17$, it is important to determine the range of redshifts and scales that future 21-cm experiments may in principle be able to access. In this paper we have addressed the following question in detail: how close to the CMB last scattering surface could 21-cm observations reach? To do so, we presented the most accurate calculation to date of the 21-cm global intensity and fluctuation amplitudes at redshifts beyond $z\sim500$. We performed a high-precision radiative transfer calculation of the spectrum near Ly$\alpha$, in order to quantify the Wouthuysen-Field coupling at high redshifts. Despite a large coupling strength due to the intense \lya background resulting from cosmological recombination, we found that the departure of the \lya color temperature from the gas temperature is substantially smaller than what was found in FL13. This is due to the very efficient frequency diffusion of photons in the core of the line, combined with a suppressed contribution of the damping wings to spin-flip scatterings. We moreover accounted for the effects of free-free processes for the first time. We found that the global 21-cm signal at  $z\gtrsim8 00$ is, in fact, determined mostly by these processes rather than the resonant interactions with the 21-cm transition. Nonetheless, 21-cm fluctuations do retain tomographic information, in principle up to  $z\sim 1100$ or so before being exponentially damped by bremsstrahlung absorption.
Finally, we estimated the detectability of the signal, accounting for smooth foreground subtraction and angular smearing due to scattering in the Milky Way's interstellar medium. We found that even a highly optimistic instrumental setup, with collecting area of the order of the Moon's surface area and a century-long observation time, could detect 21-cm fluctuations only up to redshifts of a few hundreds.

On the bright side, there still remains a spectacular trove of information to be collected with the 21-cm signal at $z \lesssim 200$ or so. As instrumentation progresses and observations push their horizon deeper in the Universe's early stages, it will be important to carefully model the expected signal. As illustrated in this work, the physics of the 21-cm signal is rich and subtle, and careful modeling could very well lead to interesting surprises.

\section*{Acknowledgements}

We thank Jens Chluba, Ely Kovetz, Marc Kamionkowski, and Anastasia Fialkov for useful conversations. Part of this work was completed at Johns Hopkins University, with funding from Simons Foundation Award 327938, NSF grant PHY-1214000, NASA ATP grant NNX15AB18G.
CH is supported by NASA, the Simons Foundation, and the US Department of Energy.

%


\appendix
\section{Numerical computation of the \lya spectrum}
\label{app:numerical}
Here we describe in detail the numerical radiative transfer computation carried out in Section \ref{sec:radtransfer}.  We use the method of HF09.  No claim of originality is made in this section, these details are included here only for the sake of completeness.

As stated above, the spectrum evolves due to Hubble drift, emissions, absorptions, and scatterings.  We consider the last of these first.  In order to conserve photons, rather than writing a derivative operator for the Hubble drift, we chose the timestep of our Fokker-Planck solver such that $\Delta\ln\nu=\Delta\ln a$.  Under this assumption, at each time step the photons in one bin simply shift to the next lower-frequency bin, i.e.
\be
N_i(a+\Delta a) = N_{i+1}(a).
\ee
We obtain the value for the bluemost bin from the HyRec spectrum.  Because we chose such a large number of bins, this bin is sufficiently far from the \lya line that the exact choice of its frequency does not affect our final results.

The emission rate of \lya photons is set by sum of the decay rates $\Gamma_{u\rightarrow2p}$ from higher-level states $u$ to the $2p$ state multiplied by the branching fraction $P_{2p\rightarrow1s}$ for $2p$ to decay to $1s$.  The contribution $\dot{N}|_{\rm{em}}$ to the time derivative of $N_i$ can be written as
\be
\dot{N}_i|_{\rm{em}}=\sum_u x_u\Gamma_{u\rightarrow2p}P_{2p\rightarrow1s}\mathcal{E}(\nu_i)\phi_\alpha(\nu_i)\Delta\nu_i,
\ee
where $x_u$ is the fraction of hydrogen atoms in state $u$.  Values for $x_{1s}$ and $x_{2p}$ are given by HyRec, populations for higher states are computed assuming thermal equilibrium.  The quantity $\mathcal{E}(\nu)$ is a correction factor for the emission rate due to photon phase space factors and the presence of other resonances, its value is given by equations (25-26) of HF09.  Once we have the emission rate, we can use detailed balance to write the absorption rate $\dot{N}_i|_{\rm{ab}}$ as
\be
\frac{\dot{N}_i|_{\rm{ab}}}{\dot{N}_i|_{\rm{em}}}=-\frac{N_i}{N_{\rm{eq}}}\left(\frac{\nu_{\LYA}}{\nu_i}\right)^3e^{h(\nu_i-\nu_{\LYA})/T_r},
\ee
where
\be
N_{\rm{eq}}=\frac{8\pi\nu_{\LYA}^3}{c^3n_{\rm{H}}}\frac{x_{2p}}{3x_{1s}}\Delta\ln\nu
\ee
is the equilibrium value of $N_i$ at line center.

True absorptions and emissions change the overall number of photons in our calculations.  We now need to add in the final effect, scattering, which preserves photon number but alters their frequency.  If $F_i$ is the net flux of photons from bin $i+1$ to $i$ due to scattering, then 
\be
\dot{N}_i|_{\rm{sc}}=F_i-F_{i-1}.
\ee
Because we are working in the Fokker-Planck approximation, these fluxes are linear in $N_i$ and $dN_i/d\nu$, so we can write
\be
F_i=-\zeta_iN_i+\eta_iN_{i+1}.
\ee
One relation between our new constants $\zeta_i$ and $\eta_i$ can be determined by requiring the $F_i$ be zero in thermal equilibrium, which yields
\be
\frac{\zeta_i}{\eta_i}=\frac{\nu_{i+1}^3}{\nu_i^3}e^{-h(\nu_{i+1}-\nu_i)/T_m}.
\label{etazeta1}
\ee
The matter temperature $T_m$ appears in the exponent because scattering effects tend to bring the color temperature in to equilibrium with the kinetic temperature of the gas.

In terms of $\zeta_i$ and $\eta_i$, the diffusion coefficient can be written as
\be
\mathcal{D}(\nu)=\frac{1}{2}(\eta_i+\zeta_{i-1})(\Delta\nu)^2.
\ee
Comparing this to the actual value of $\mathcal{D}(\nu)$ from \citet{Hirata2006} and assuming that $\zeta_i$ and $\eta_i$ are slowly varying functions yields the result that
\be
\eta_i+\zeta_i=\frac{H\nu_{\LYA}\sigma_\nu^2\tau_{\LYA}\left[\phi_\alpha(\nu_i)+\phi_\alpha(\nu_{i+1})\right]}{(\nu_{i+1}-\nu_i)^2},
\label{etazeta2}
\ee
where $\tau_{\LYA}$ is the optical depth in the \lya line.  By combining equations (\ref{etazeta1}) and (\ref{etazeta2}), we can determine $F_i$ and therefore $\dot{N}_i|_{\rm{sc}}$.  Note that in this expression we have set the fraction of absorptions which lead to scattering equal to unity, which should be true to high precision.  We assume that there is no scattering on either boundary of our calculation.  

We solve our Fokker-Planck problem using a backwards Euler solver.  We can write the full time derivative of $N_i$ as
\be
\dot{N}_i=\dot{N}_i|_{\rm{em}}+\dot{N}_i|_{\rm{ab}}+\dot{N}_i|_{\rm{sc}}=C_{ij}N_j.
\label{tridiag}
\ee
The matrix $C_{ij}$ is tridiagonal, since the derivative in one bin depends only on the values in itself and those immediately adjacent.  On each timestep, we solve 
\be
\frac{N_i(t+\Delta t)-N_i(t)}{\Delta t}=C_{ij}(t+\Delta t)N_j(t+\Delta t)
\ee
using the standard tridiagonal method to obtain new values of $N_i$.  We then shift each value redward by one bin, filling in the bluemost bin with the value obtained from \textsc{HyRec}.  


\section{Convergence Tests}
\label{app:convergence}
The 21 cm signal we predict here is the result of a sub-percent difference between the spin and radiation temperatures.  It is therefore crucial that we make certain that our numerical calculations are as accurate as possible.  The most likely place for numerical errors to creep in is the \lya radiative transfer computation described in Section \ref{sec:radtransfer}.  In this appendix, we will briefly describe the efforts we have made to demonstrate the convergence of this code.

We mentioned above that we use nearly 100 times higher spectral and temporal resolution for our code than was used in the similar work by HF09.  This is required to ensure that the spectrum in both the core and wings of the \lya line converges to the precision we require.  If we use the bin sizes specified in HF09, we find that amplitude of the plateau in $1-T_C/T_r$ from Figure \ref{fig:Tcolor} is higher by roughly a factor of two.  In the lower resolution calculation, the slope of the blue side spectrum is slightly steeper than that plotted in Figure \ref{fig:spectrum}.  This drives the color temperature further out of equilibrium than it otherwise should be.  It is unclear whether this discrepancy is sourced by insufficient temporal resolution or insufficient spectral resolution, as our prescription for Hubble redshifting requires us to keep the two step sizes equal.  We selected the final resolution used in the body of the paper by gradually increasing the resolution until the curves plotted in Figure \ref{fig:Tcolor} converged.  At our full resolution, halving the time and frequency steps changes the values of those curves by $<$1\%.

Similarly, we use nearly 100 times more frequency bins than HF09 in our calculation.  At fixed frequency step, the number of bins sets the total width in frequency space of our high-precision computation.  Because the wings of the \lya transition allow for significant scattering of photons well outside the line center, it is necessary to use quite wide boundary conditions.  With the HF09 bin counts, small differences between the \textsc{HyRec} spectrum and the correct value contaminate our color temperature.  We selected our final bin counts using the same criterion as for the spectral resolution.

It is also possible that there could be some error introduced by our choice of high-redshift boundary condition.  We use a blackbody initial spectrum, but is unlikely that the spectrum at our initial redshift is an exact blackbody.  However, because the optical depth to \lya scatterings is so high, any effects caused by this choice are quickly damped away.  For example, if we start our calculation at $z=1500$ instead of 1600 the value of $1-T_C/T_r$ converges to that plotted in Figure \ref{fig:Tcolor} to within $<$0.1\% by $z=1490$.  We can therefore be fairly certain that the results presented above are independent of our choice of initial condition. 


\end{document}